\documentclass[nobibnotes,preprintnumbers,aps,prl,superscriptaddress,showpacs,twocolumn,twoside,sort&compress]{revtex4}				 
\usepackage{graphicx}																												 
\usepackage{titlesec}																											     
\usepackage[colorlinks=true,linkcolor=blue,citecolor=red,urlcolor=blue]{hyperref}													 

\begin{document}																													 
%
																												

\newcommand{\kvec}{\mbox{{\scriptsize {\bf k}}}}
\newcommand{\lvec}{\mbox{{\scriptsize {\bf l}}}}
\newcommand{\qvec}{\mbox{{\scriptsize {\bf q}}}}

\def\eq#1{(\ref{#1})}
\def\fig#1{\hspace{1mm}Fig. \ref{#1}}
\def\tab#1{\hspace{1mm}Table \ref{#1}}
\title{Quantitative analysis of nonadiabatic effects in dense H$_3$S and PH$_3$ superconductors
}
\author{Artur P. Durajski} \email{adurajski@wip.pcz.pl}
\affiliation{Institute of Physics, Cz{\c{e}}stochowa University of Technology, Ave. Armii Krajowej 19, 42-200 Cz{\c{e}}stochowa, Poland}
\date{\today} 

\begin{abstract}
The comparison study of high pressure superconducting state of recently synthesized H$_3$S and PH$_3$ compounds are conducted within the framework of the strong-coupling theory.
By generalization of the standard Eliashberg equations to include the lowest-order vertex correction, we have investigated the influence of the nonadiabatic effects on the Coulomb pseudopotential, electron effective mass, energy gap function and on the $2\Delta(0)/T_C$ ratio. 
We found that, for a fixed value of critical temperature ($178$ K for H$_3$S and $81$ K for PH$_3$), the nonadiabatic corrections reduce the Coulomb pseudopotential for H$_3$S from $0.204$ to $0.185$ and for PH$_3$ from $0.088$ to $0.083$, however, the electron effective mass and ratio $2\Delta(0)/T_C$ remain unaffected.
Independently of the assumed method of analysis, the thermodynamic parameters of superconducting H$_3$S and PH$_3$ strongly deviate from the prediction of BCS theory due to the  strong-coupling and retardation effects.
\\

\end{abstract}


\maketitle
%
\section{Introduction}
%
The first-principles theoretical studies of the metallization and high-temperature superconductivity of dense hydrogen sulfide were reported for the first time by Li \textit{et al.} in 2014 \cite{Li2014A}. 
This directly initiated the experimental work of Drozdov \textit{et al.} who found that H$_2$S compressed
in a diamond anvil cell exhibit the superconductivity ranging from $30$ to $150$ K measured in the low-temperature runs \cite{Drozdov2015A, YanmingMa}, which is consistent with calculations mentioned \cite{Li2014A}. Furthermore, the results achieved in samples prepared at high-temperature showed that the record critical temperature of $164$ K for cooper-oxide system $\rm HgBa_{2}Ca_2Cu_3O_{8+\delta}$ under quasihydrostatic pressure \cite{GaoHg} has been trumped.
Based on a sharp drop of the resistivity to zero and an expulsion of the magnetic field, Drozdov \textit{et al.} observed a transition from metal to superconducting state at $203$ K in H$_2$S sample compressed up to $155$ GPa \cite{Drozdov2015A, Troyan2016}. 
Subsequent theoretical \cite{Bernstein, Duan180502, Errea2015A} and experimental \cite{Li020103, Einaga2016A}  studies suggested that at high pressure, the phase diagram favors decomposition of H$_2$S into H$_3$S and elemental sulfur. 
This result means that the superconducting state observed at $203$ K comes from a decomposition product H$_3$S.
More recently, referring to the theoretical crystal structure searches performed by Duan \textit{et al.} \cite{Duan2014A}, the stability of high-pressure cubic $Im\overline{3}m$ structure of H$_3$S was confirmed 
by Li \textit{et al.} in first-principles DFT structure searches joined with high-pressure X-ray diffraction experiments \cite{Li020103}, and then by Einaga \textit{et al.} in synchrotron X-ray diffraction measurements combined with the electrical resistance measurements \cite{Einaga2016A}.
In contrast to cuprates where the nature of superconductivity is still not fully understood \cite{GZhao, Szczesniak2012PLOS, Chen2014}, the presence of a strong isotope effect in H$_3$S clearly suggests the electron--phonon origin of the superconducting state \cite{Drozdov2015A, Mazin, Nicol, Ortenzi}.
In addition, Jarlborg and Bianconi predict that the Fermi surface of H$_3$S consists of multiple sheets similar to those in MgB$_2$ \cite{Bianconi1}. However, the existence of multi-gap superconductivity in H$_3$S has not been so far confirmed.

\begin{figure}[!h]
\includegraphics[width=\columnwidth]{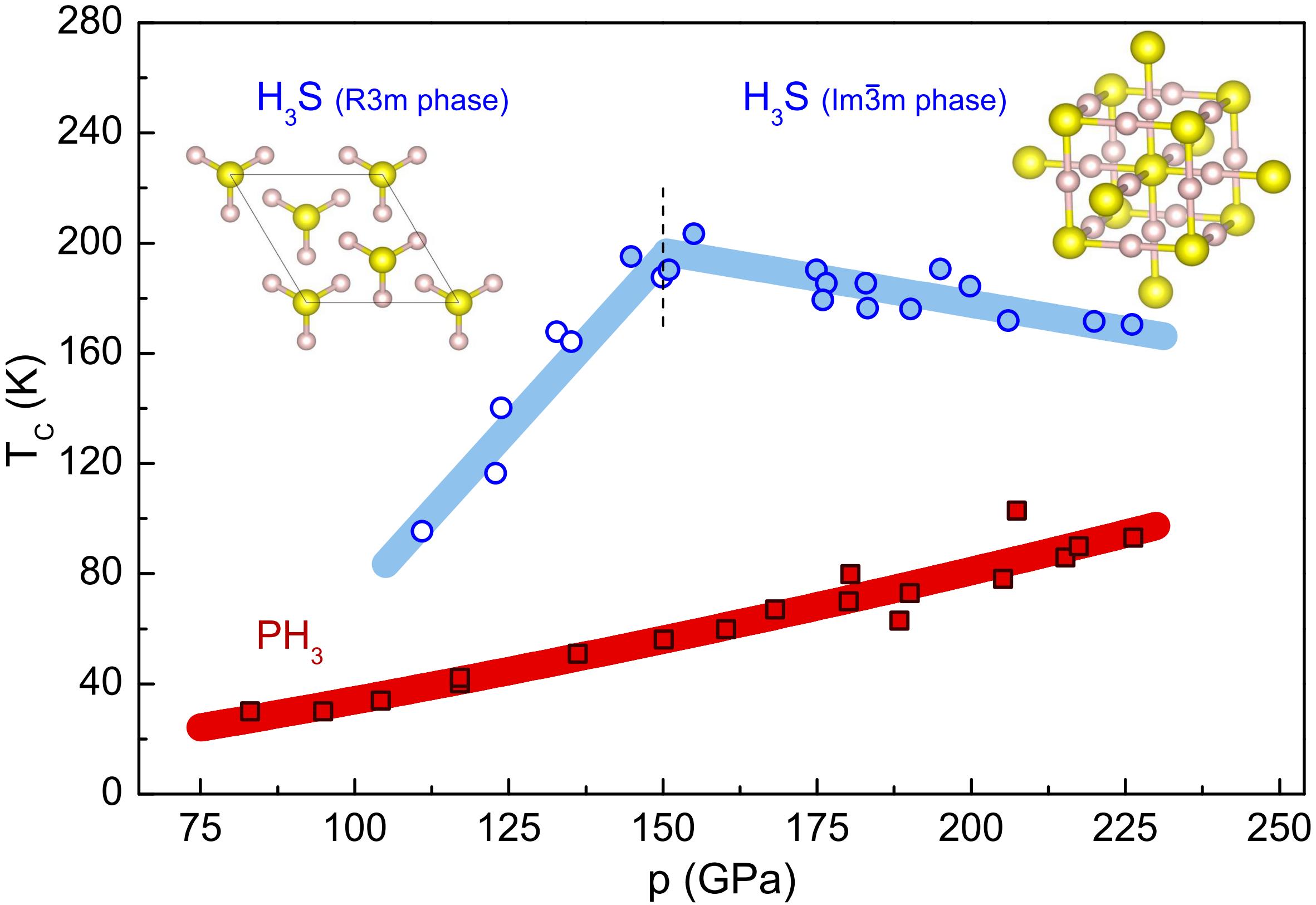}
\caption{The experimental data of critical temperature as a function of pressure for H$_3$S \cite{Drozdov2015A, Einaga2016A} and PH$_3$ \cite{Drozdov2015PH3}. In addition the stable structures of H$_3$S \cite{Duan2014A} are presented.}
\label{f01}
\end{figure}

The above theoretical and experimental discovery have stimulated significant interest in finding new hydrogen-containing superconductors \cite{GeYanfeng, ZhongXin, ShoutaoZhang}. Very recently, Drozdov \textit{et al.} reported superconductivity in compressed PH$_3$ with a $T_C$ above $100$ K \cite{Drozdov2015PH3}.
The pressure dependence of the experimental critical temperature for H$_3$S and PH$_3$ compounds together with the relevant crystal structures is presented in \fig{f01}.
For H$_3$S the second-order structural phase transition from trigonal $R3m$ to cubic $Im\overline{3}m$ is experimentally observed at pressure close to $150$ GPa \cite{Einaga2016A, Gorkov}.
However, according to the static lattice calculations the phase transition from $R3m$ to $Im\overline{3}m$ occurs at $\sim180$ GPa \cite{Duan2014A}. A recent theoretical study found that quantum nuclear motion lowers the transition pressure to $103$ GPa \cite{Errea2016A}.
In the case of PH$_3$, a theoretical search of crystal structure reveals two phases with lowest energy: orthorhombic $P2_12_12_1$ and monoclinic $C2/m$. 
The DFT studies realized by Liu \textit{et al.} indicate that both structures are dynamically stable and superconducting but $C2/m$ phase are in a better agreement with an experimental results \cite{LiuPH3}.
Unfortunately, the lack of structural informations on the superconducting phases from suitable measurements do not allow at this moment for the unambiguous verification of these assumptions.

Motivated by a recent significant experimental and theoretical progress in chemistry and physics of hydrogen-dense materials, we have carried out calculations to explore in detail the thermodynamic properties of superconducting hydrogen sulfide H$_3$S and hydrogen phosphide PH$_3$ at extremely high pressure ($p=200$ GPa).
The very large values of electron-phonon coupling interaction observed in these systems, caused that our investigations were performed within the framework of the Migdal-Eliashberg (ME) theory of superconductivity \cite{Eliashberg1960A}, which goes beyond the BCS model \cite{Bardeen1957B} by taking into account the retardation and strong-coupling effects.

This paper is organized as follows. In Section II, we introduce the theoretical model used to determine the quantities characterizing the superconducting state. Moreover, we present details of the first-principles calculations carried out to study the phonon properties and electron-phonon interactions.
Then, in Section III, we report and compare the thermodynamic properties of superconducting H$_3$S and PH$_3$ systems at $200$ GPa. 
We discuss the validity of the conventional Migdal-Eliashberg theory by introduce the lowest-order vertex correction and we examine its effect on Coulomb pseudopotential, energy gap, $2\Delta(0)/T_C$ ratio and electron effective mass.
Finally, we give a summary of this study in Section IV.

\section{Theoretical model and computational methods}

%
\begin{figure*}
\includegraphics[width=1.2\columnwidth]{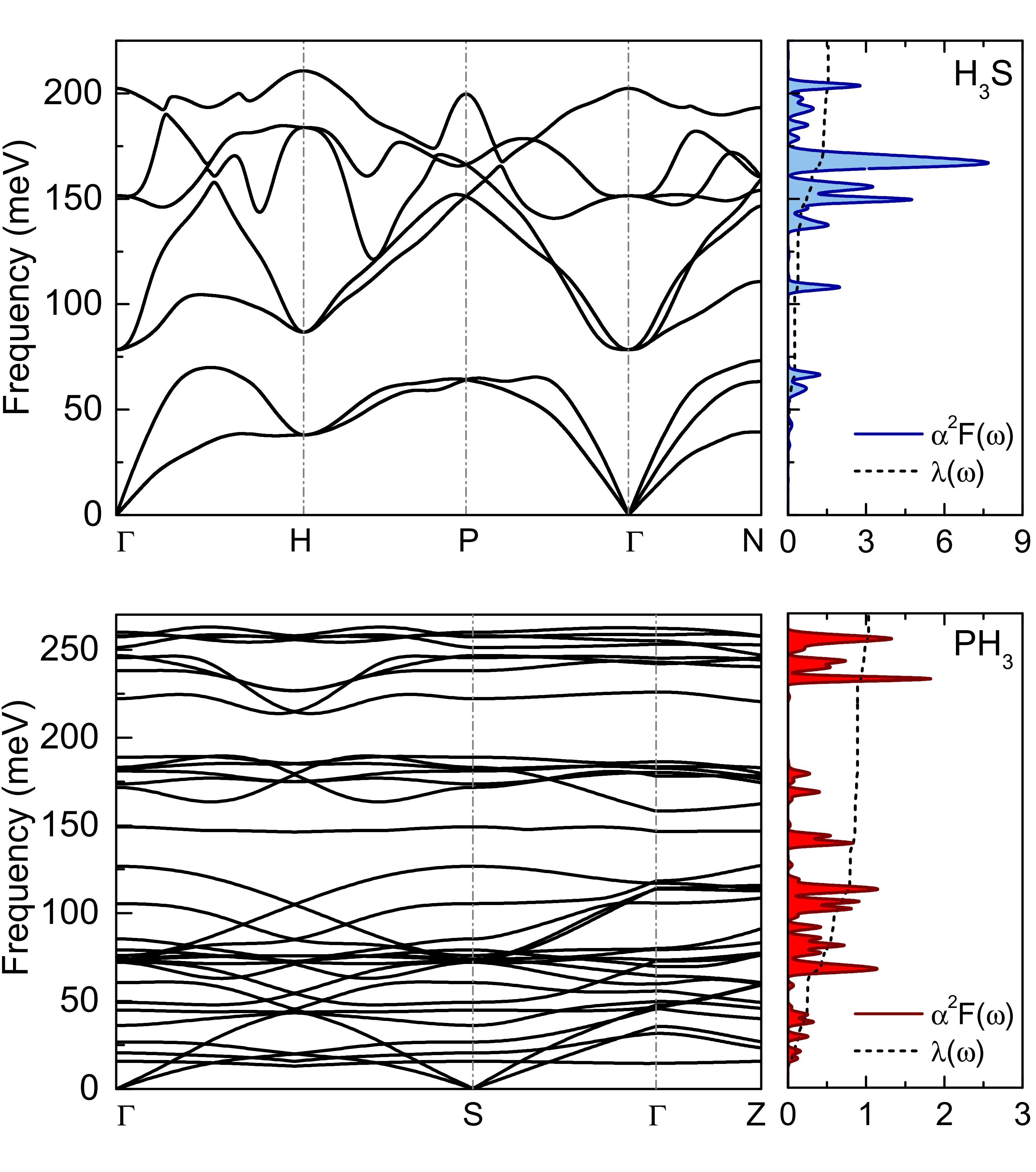}
\caption{Phonon dispersion and the Eliashberg spectral function $\alpha^{2}F(\omega)$ with electron-phonon integral $\lambda(\omega)$ for H$_3$S and PH$_3$ superconductors at $200$ GPa.}
\label{f2}
\end{figure*}
%
Besides the experimental works, most of the theoretical studies concluded that compressed H$_3$S and PH$_3$ are phonon-mediated strong-coupling superconductors \cite{Duan2014A, LiuPH3, AkashiH2S, Errea2015A, FloresL2016}. Thus, the superconducting state of these compounds can be accurately described by the Migdal-Eliashberg theory \cite{Durajski2015A, Durajski2015B}.
The Eliashberg equations for the superconducting order parameter function $\varphi_{n}\equiv\varphi\left(i\omega_{n}\right)$ and the electron mass renormalization function $Z_{n}\equiv Z\left(i\omega_{n}\right)$ written in the imaginary-axis formulation are given by \cite{Eliashberg1960A, Marsiglio1988A}:
\begin{equation}
\label{r1}
\varphi_{n}=\frac{\pi}{\beta}\sum_{m=-M}^{M}
\frac{\lambda_{n,m}-\mu^{\star}\theta\left(\omega_{c}-|\omega_{m}|\right)}
{\sqrt{\omega_m^2Z^{2}_{m}+\varphi^{2}_{m}}}\varphi_{m},
\end{equation}
and
\begin{equation}
\label{r2}
Z_{n}=1+\frac{1}{\omega_{n}}\frac{\pi}{\beta}\sum_{m=-M}^{M}
\frac{\lambda_{n,m}}{\sqrt{\omega_m^2Z^{2}_{m}+\varphi^{2}_{m}}}
\omega_{m}Z_{m},
\end{equation}
where the pairing kernel for the electron-phonon interaction is given by:
\begin{equation}
\label{r3}
\lambda_{n,m}= 2\int_0^{\omega_{{D}}}d\omega\frac{\omega}
{\left(\omega_n-\omega_m\right)^2+\omega ^2}\alpha^{2}F\left(\omega\right).
\end{equation}
Moreover, $\beta=1/k_BT$ and $k_B=0.0862$ meV/K states the Boltzmann constant. 
Symbols $\mu^{\star}$ and $\theta$ denote the screened Coulomb repulsion and the Heaviside function with cut-off frequency $\omega_c$ equal to ten times the maximum phonon frequency: $\omega_c=10\omega_D$.

The application of the above Eliashberg equations to describe the electron-phonon superconductivity is justified for systems in which the value of the phonon energy scale (Debye frequency, $\omega_D$) to the electron energy scale (Fermi energy, $\varepsilon_F$) ratio is negligible.
Otherwise the Eliashberg equations should be generalized by taking into account the lowest-order vertex correction \cite{Freericks1998A, Pietronero, Grimaldi1999, GrimaldiPietronero, PietroneroEPL}:
\begin{widetext}
\begin{eqnarray}
\label{r1}
\varphi_{n}&=&\pi T\sum_{m=-M}^{M}
\frac{\lambda_{n,m}-\mu^{\star}\theta\left(\omega_{c}-|\omega_{m}|\right)}
{\sqrt{\omega_m^2Z^{2}_{m}+\varphi^{2}_{m}}}\varphi_{m}\\ \nonumber
&-&
\frac{\pi^{3}T^{2}}{4\varepsilon_{F}}\sum_{m=-M}^{M}\sum_{m'=-M}^{M}
\frac{\lambda_{n,m}\lambda_{n,m'}}
{\sqrt{\left(\omega_m^2Z^{2}_{m}+\varphi^{2}_{m}\right)
       \left(\omega_{m'}^2Z^{2}_{m'}+\varphi^{2}_{m'}\right)
       \left(\omega_{-n+m+m'}^2Z^{2}_{-n+m+m'}+\varphi^{2}_{-n+m+m'}\right)}}\\ \nonumber
&\times&
\left[
\varphi_{m}\varphi_{m'}\varphi_{-n+m+m'}+2\varphi_{m}\omega_{m'}Z_{m'}\omega_{-n+m+m'}Z_{-n+m+m'}-\omega_{m}Z_{m}\omega_{m'}Z_{m'}
\varphi_{-n+m+m'}
\right],
\end{eqnarray}
and
\begin{eqnarray}
\label{r2}
Z_{n}&=&1+\frac{\pi T}{\omega_{n}}\sum_{m=-M}^{M}
\frac{\lambda_{n,m}}{\sqrt{\omega_m^2Z^{2}_{m}+\varphi^{2}_{m}}}\omega_{m}Z_{m}\\ \nonumber
&-&
\frac{\pi^{3}T^{2}}{4\varepsilon_{F}\omega_{n}}\sum_{m=-M}^{M}\sum_{m'=-M}^{M}
\frac{\lambda_{n,m}\lambda_{n,m'}}
{\sqrt{\left(\omega_m^2Z^{2}_{m}+\varphi^{2}_{m}\right)
       \left(\omega_{m'}^2Z^{2}_{m'}+\varphi^{2}_{m'}\right)
       \left(\omega_{-n+m+m'}^2Z^{2}_{-n+m+m'}+\varphi^{2}_{-n+m+m'}\right)}}\\ \nonumber
&\times&
\left[
\omega_{m}Z_{m}\omega_{m'}Z_{m'}\omega_{-n+m+m'}Z_{-n+m+m'}+2\omega_{m}Z_{m}\varphi_{m'}\varphi_{-n+m+m'}-\varphi_{m}\varphi_{m'}\omega_{-n+m+m'}Z_{-n+m+m'}
\right],
\end{eqnarray}
\\
where, the modified electron-phonon pairing kernel takes the following form:

\begin{eqnarray}
\label{r3}
\lambda_{n,m}=2\int_0^{\omega_{D}}d\omega\frac{\omega}{\omega ^2+4\pi^{2}T^{2}\left(\omega_n-\omega_m\right)^{2}}\alpha^{2}F\left(\omega\right).
\end{eqnarray}
\end{widetext}

The Eliashberg spectral function, one of the main input element to the Eliashberg equations, is defined as:
\begin{equation}
\alpha^2F(\omega) = {1\over 2\pi N(0)}\sum_{{\bf q}\nu} 
                    \delta(\omega-\omega_{{\bf q}\nu})
                    {\gamma_{{\bf q}\nu}\over\hbar\omega_{{\bf q}\nu}}
\end{equation}

with
\begin{eqnarray}
\gamma_{{\bf q}\nu} &=& 2\pi\omega_{{\bf q}\nu} \sum_{ij}
                \int {d^3k\over \Omega_{BZ}}  |g_{{\bf q}\nu}({\bf k},i,j)|^2
                    \delta(\epsilon_{{\bf q},i} - \epsilon_F) \\\nonumber  &\times& \delta(\epsilon_{{\bf k}+{\bf q},j} - \epsilon_F), 
\end{eqnarray}
where $N(0)$, $\gamma_{{\bf q}\nu}$ and $g_{{\bf q}\nu}({\bf k},i,j)$ denote the density of states at the Fermi energy, the phonon linewidth and the electron-phonon coefficients, respectively.
The $\alpha^{2}F(\omega)$ functions for H$_3$S and PH$_3$ were calculated in this paper using density functional perturbation theory and the plane-wave pseudopotential method, as implemented in the Quantum-Espresso package \cite{QE}. 
We assume that H$_3$S has cubic crystal $Im\overline{3}m$ structure with lattice parameter $a=2.984$ $\rm \AA$ \cite{Duan2014A}, whereas PH$_3$ crystallize in monoclinic $C2/m$ structure with lattice parameter $a=5.152$ $\rm \AA$, $b=2.961$ $\rm \AA$, $c=2.960$ $\rm \AA$, $\alpha=\gamma=90^{\circ}$ and $\beta=90.23^{\circ}$ \cite{Shamp2016}.
The Vanderbilt-type ultrasoft pseudopotentials for S, P and H atoms were employed with a kinetic energy cutoff equal to $80$ Ry. The phonon calculations were performed for $\rm 32\times 32\times 32$ Monkhorst-Pack $k$-mesh with gaussian smearing of $0.03$ Ry. The electron-phonon coupling matrices were computed using $\rm 8\times 8\times 8$ $q$-grid for H$_3$S and $\rm 4\times 4\times 4$ $q$-grid for PH$_3$.
The calculated phonon band structures together with the Eliashberg spectral functions and electron-phonon integrals $\lambda(\omega)=2\int_0^{\Omega_{\rm{max}}}{\omega^{-1}}{\alpha^2F(\omega)}d\omega$ for both investigated hydrides are presented in \fig{f2}.
The absence of imaginary frequencies in the full phonon spectra indicates that both systems are dynamically stable.

%
\begin{figure}[!h]
\includegraphics[width=\columnwidth]{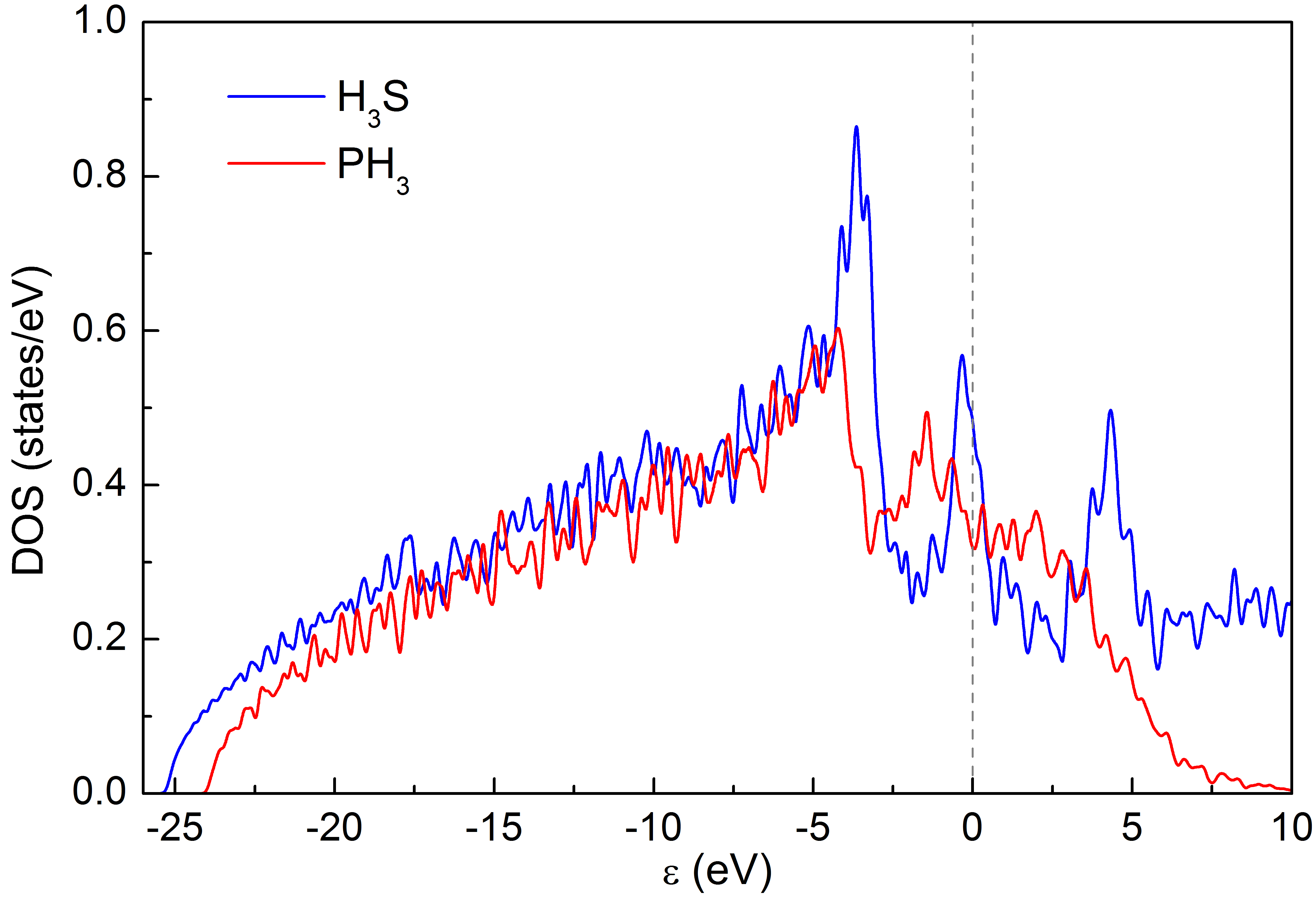}
\caption{Density of states (DOS) for H$_3$S ($Im\overline{3}m$) and PH$_3$ ($C2/m$) at $200$ GPa. The dotted line at zero indicates the Fermi level.}
\label{f2a}
\end{figure}
%
The electronic band structures of H$_3$S and PH$_3$ at $200$ GPa were also explored. The investigated compounds are good metals with a large density of states (DOS) at the Fermi level. To clarify the difference of the electronic structures between $Im\overline{3}m$ H$_3$S and $C2/m$ PH$_3$, the DOS calculated for these two crystal structures are shown in \fig{f2a}. A strong peak around the Fermi level (the Van Hove singularities) is favorable for a strong electron-phonon coupling and thus really high superconducting temperature in the case of H$_3$S \cite{Sano}.

Our \textit{ab-initio} studies showed that ratio $\lambda\omega_D/\varepsilon_F$ is equal to $0.020$ for H$_3$S and $0.014$ for PH$_3$. 
These values are rather small in comparison to fullerene compounds or high-$T_c$ cuprates, however, are not zero.
Due to the above, we decided to conduct our calculations simultaneously using the conventional Eliashberg equations and equations with the lowest-order vertex correction, which allows us to examine the influence of nonadiabatic effects on the thermodynamic properties in studied compounds.
The numerical analysis was performed using a self-consistent iteration methods \cite{szczesniak2006A}, which were implemented successfully in our previous papers \cite{Durajski2015B, Szczesniak2015A, Szczesniak2015SSC, Szczesniak2016PhysB}.
The convergence and precision of our results are controlled by assuming the sufficiently high number ($M=1100$) of Matsubara frequencies $\omega_{n}\equiv\left(\pi/\beta\right)\left(2n-1\right)$, where $n=0,\pm 1,\pm 2,\dots,\pm M$.

\section{Results and discussion}

\begin{figure}[!h]
\includegraphics[width=\columnwidth]{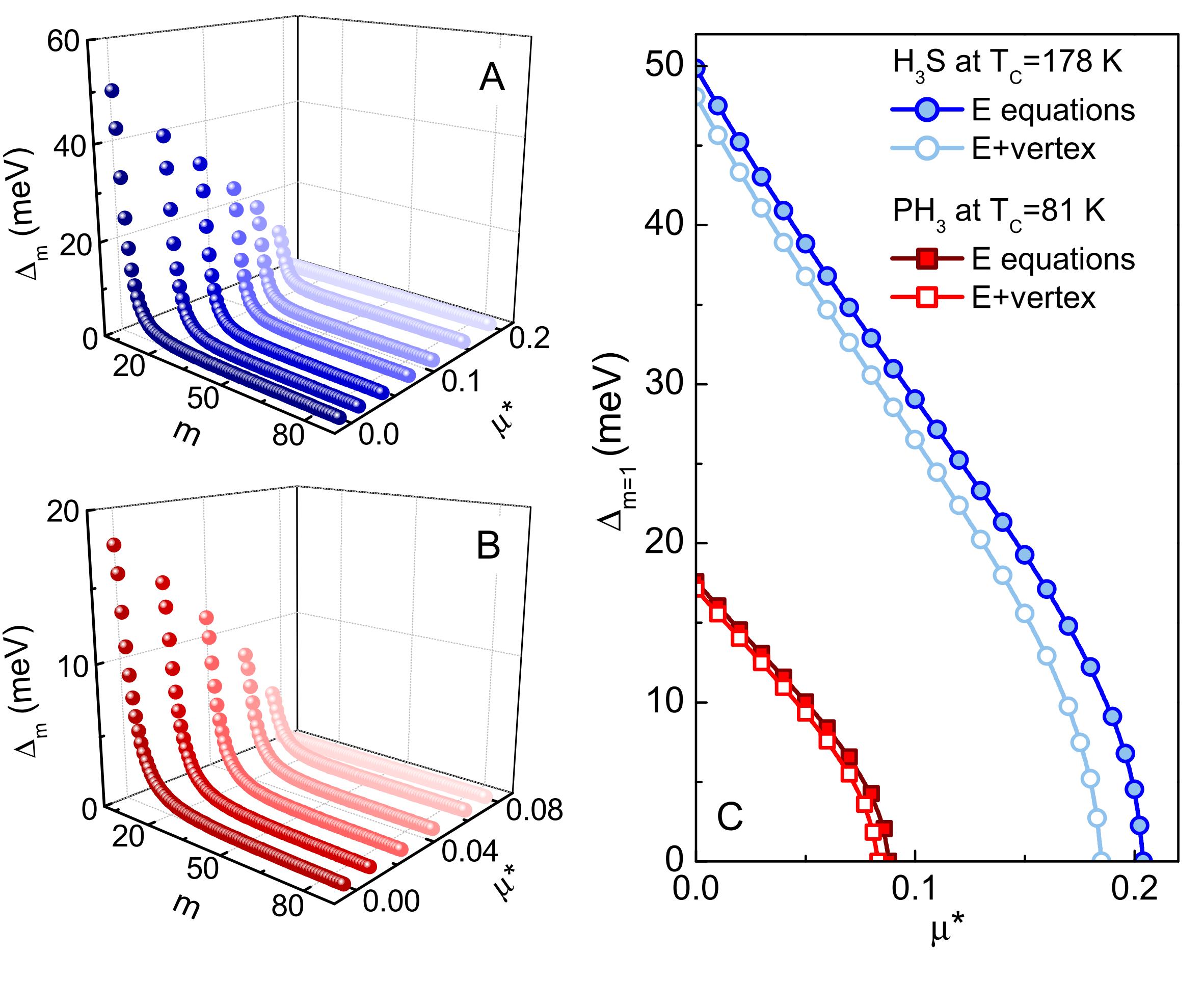}
\caption{The order parameter on the imaginary axis as a function of $m$ and $\mu^{\star}$ for (A) H$_3$S and (B) PH$_3$ at $200$ GPa - the results was obtained using conventional Eliashberg (E) equations. (C) The full dependence of the first value of the order parameter as a function of Coulomb pseudopotential.}
\label{f3}
\end{figure}

To study the thermodynamic properties of phonon-mediated superconductors on the quantitatively level in the first step we determine the critical value of the Coulomb pseudopotential $\mu^{\star}_C$.
For this purpose, in the Eliashberg equations we replace $T$ by the experimental value of critical temperature: $T_C=178$ K for H$_3$S \cite{Drozdov2015A} and $T_C=81$ K for PH$_3$ \cite{Drozdov2015PH3}.
Then, starting from zero, we increase the value of $\mu^{\star}$ until we reach the equality $\Delta_{m=1}(\mu^{\star}=\mu^{\star}_C)=0$, where the order parameter is defined as: $\Delta_{m}=\varphi_{m}/Z_{m}$ \cite{Carbotte1990A}. The obtained results are presented in \fig{f3}.
%
%
In particular, on the basis of the full dependence of $\Delta_{m=1}$($\mu^{\star}$) we can conclude that if we take into account the conventional Eliashberg equations, the Coulomb pseudopotential takes a relatively high critical value for H$_3$S ($\mu^{\star}_C=0.204$) and low value for PH$_3$ ($\mu^{\star}_C=0.088$) at $200$ GPa. 
Moreover, at this point we can found that, for a fixed value of critical temperature, the lowest-order vertex correction changes $\mu^{\star}_C$ by $-9.3\%$ for H$_3$S and $-5.7\%$ for PH$_3$. 

\begin{figure}[!h]
\includegraphics[width=\columnwidth]{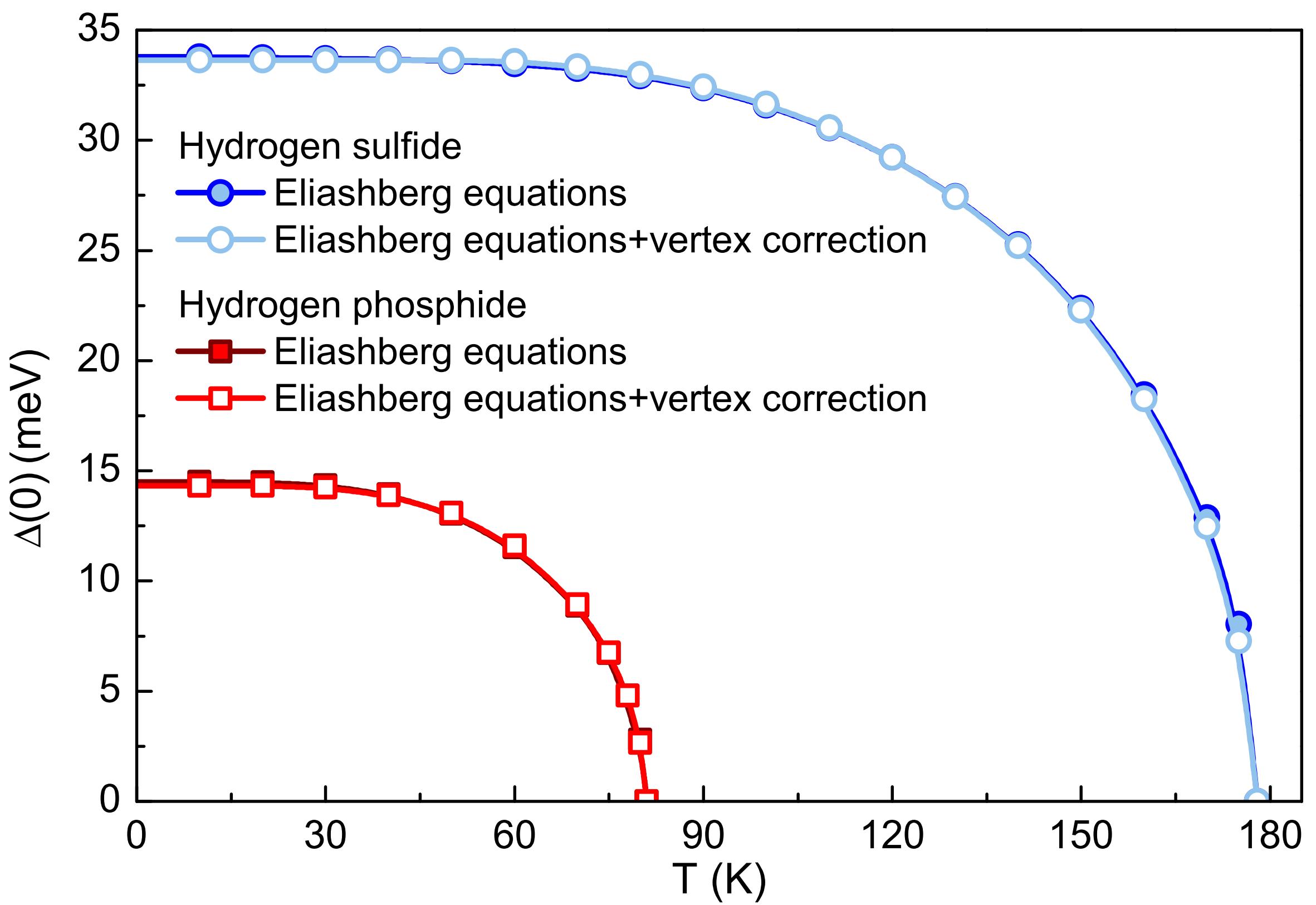}
\caption{The temperature dependence of the superconducting energy gap. The numerical results can be reproduced using the analytical formula $\Delta\left(T\right)=\Delta\left(0\right)\sqrt{1-\left(T\slash T_{C}\right)^{\alpha}}$, where $\alpha=3.36$.}
\label{f4}
\end{figure}

In the next step, by using the analytical continuation of the imaginary-axis solutions to the real frequency axis \cite{AnalyticContinuation}, we calculate the temperature dependence of superconducting energy gap $\Delta\left(T\right)={\rm Re}\left[\Delta\left(\omega=\Delta\left(T\right),T\right)\right]$ \cite{Carbotte1990A}. The obtained results are presented in \fig{f4}. 
%
%
Knowledge of the energy gap value at zero temperature allowed us to calculate the dimensionless ratio $R_{\Delta}\equiv2\Delta(0)/T_{C}$ for which the BCS theory predicts universal value $[R_{\Delta}]_{\rm BCS}=3.53$.
In the case of studied hydrides, $R_{\Delta}$ exceed the value of $4$, in particular by using the conventional Eliashberg equations we received $[R_{\Delta}]_{\rm H_3S}=4.41$ and $[R_{\Delta}]_{\rm PH_3}=4.15$.
This behavior is connected with the strong-coupling and retardation effects, which in the framework of the Eliashberg formalism can be characterized by the ratio $T_C/\omega_{\rm ln}$, where $\omega_{\rm ln}$ is the logarithmic phonon frequency, correspondingly $131$ meV for H$_3$S and $79$ meV for PH$_3$. Thus, the considered ratio equals $0.12$ and $0.09$, respectively, while in the weak-coupling BCS limit we have: $T_{C}/\omega_{\rm ln}\rightarrow 0$.
Moreover, we find that although the vertex corrections seriously reduce $\mu_C^{\star}$ for H$_3$S and PH$_3$, the ratio $2\Delta(0)/T_{C}$ remains practically unaffected.
Similar situation is observed in the case of electron effective mass at $T_C$ calculated from $m_{e}^{\star}={\rm Re}\left[Z\left(0\right)\right]m_{e}$, where $m_{e}$ denotes the electron band mass. In our case $m_{e}^{\star}/m_{e}=2.736$ for H$_3$S and $m_{e}^{\star}/m_{e}=2.136$ for PH$_3$, regardless of the equations applied.
This means that the thermodynamic properties of phonon-mediated superconductors can be successfully obtained in the framework of the conventional Migdal-Eliashberg formalism with the proviso that $\mu_C^{\star}$ has to be accurately determined.

\begin{figure}[!h]
\includegraphics[width=\columnwidth]{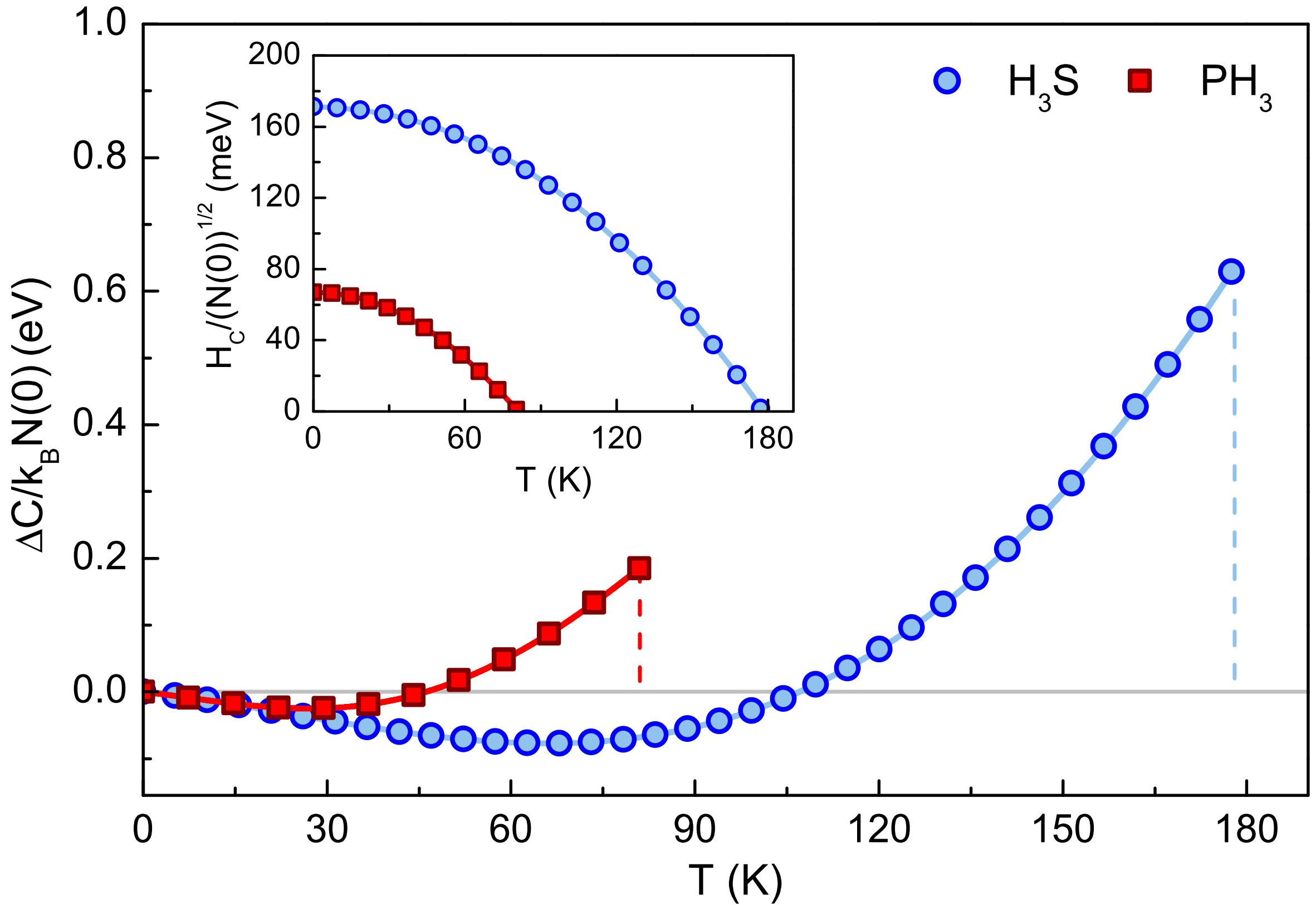}
\caption{The specific heat difference between the superconducting and the normal state as a function of temperature.
Inset presents the temperature dependence of the thermodynamic critical field.}
\label{f5}
\end{figure}

In the last step, to investigate the specific heat and thermodynamic critical field behavior, the condensation energy was numerically calculated:
\begin{eqnarray}
\label{rF}
\frac{E_{\rm cond}(T)}{N(0)}&=&-\frac{2\pi}{\beta}\sum_{n=1}^{M}
\left(\sqrt{\omega^{2}_{n}+\Delta^{2}_{n}}- \left|\omega_{n}\right|\right)\\ \nonumber
&\times&\left(Z^{N}_{n}\frac{\left|\omega_{n}\right|}
{\sqrt{\omega^{2}_{n}+\Delta^{2}_{n}}}-Z^{S}_{n}\right),
\end{eqnarray}  
where $Z^{N}_{n}$ and $Z^{S}_{n}$ denote the mass renormalization functions for the normal and superconducting states, respectively.
The specific heat difference between superconducting and normal state was then obtained from the second derivative of $E_{\rm cond}(T)$:
\begin{equation}
\label{rC}
\frac{\Delta C (T)}{k_{B}N(0)}=\frac{1}{\beta}\frac{d^{2}\left[E_{\rm cond}(T)/N(0)\right]}{d\left(k_{B}T\right)^{2}}
\end{equation}
and thermodynamic critical field is defined as:
%
\begin{equation}
\label{rH}
\frac{H_{C}(T)}{\sqrt{N(0)}}=\sqrt{8\pi\left[E_{\rm cond}(T)/N(0)\right]}.
\end{equation}
The temperature dependence of $\Delta C$, with characteristic specific heat jump at $T_C$ marked by vertical line, is presented in \fig{f5}.
%

%
The inset shows the thermodynamic critical field for investigated hydrides.
These results allow us to determine the other two fundamental dimensionless ratios: $R_{H}\equiv{T_{C}C^{N}\left(T_{C}\right)}/{H_{C}^{2}\left(0\right)}$
and $R_{C}\equiv{\Delta C\left(T_{C}\right)}/{C^{N}\left(T_{C}\right)}$, where the specific heat for the normal state is defined as $C^{N}=\gamma T$, and $\gamma$ denotes the Sommerfeld constant: $\gamma\equiv ({2}/{3})\pi^{2}\left(1+\lambda\right)k_{B}^2N(0)$.
It is noteworthy that, in the framework of the BCS theory, these ratios adopt universal values of $0.168$ and $1.43$, respectively \cite{Carbotte1990A}.
We emphasize that these dimensionless ratios, similar to that of energy gap, take non-BCS values, in particular: $R_{H}=0.136$, $R_{C}=2.47$ for H$_3$S and $R_{H}=0.150$, $R_{C}=1.99$ for PH$_3$.
We can see that with increasing values of $T_C/\omega_{ln}$ ratio the thermodynamic properties take increasingly non-BCS behavior.
%
\section{Conclusions}
In this work, using the first-principles calculations and Eliashberg theory with and without vertex corrections, we systematically study the nonadiabatic effects on the superconductivity of compressed H$_3$S and PH$_3$ compounds. We find that for a fixed experimental value of critical temperature the lowest-order vertex correction reduces the Coulomb pseudopotential (from $0.204$ to $0.185$ in the case of H$_3$S and from $0.088$ to $0.083$ for PH$_3$), however, the energy gap, electron effective mass and ratio $2\Delta(0)/T_{C}$ remain unaffected. 
It means that the superconducting behavior can be properly determined even in the framework of the classical Migdal-Eliashberg formalism, as long as the value of the Coulomb pseudopotential is correctly determined.
Moreover, we calculated the specific heat and thermodynamic critical field and we proved that strong-coupling and retardation effects caused that thermodynamic properties of H$_3$S and PH$_3$ at high pressures cannot be correctly estimated in the framework of the BCS theory.
%

%
\section{Acknowledgements}
Author would like to express his gratitude to Krzysztof Durajski, for his assistance in the numerical calculations.
Research reported in this paper was financially supported by the Cz{\c{e}}stochowa University of Technology under Grant No. BS/MN-203-301/2016.
%

%


%

\end{document}